\def \yskip{\penalty-50\vskip3pt plus 3pt minus 2pt}
\def \reference{\par \yskip \noindent \hangindent .4in \hangafter 1}
\def \abc#1#2#3#4 {\reference#1, {\sl#2}, {\bf#3}, #4}
\def \blank {\lower 5pt\hbox to 0.75in{\hrulefill}}

\def \cm{~\rm{cm}}
\def \s{~\rm{s}}
\def \km{~\rm{km}}

\def \AU{~\rm{AU}}

\def \yrs{~\rm{yrs}}
\def \yr{~\rm{yr}}

\def \lesssim{\mathrel{<\kern-1.0em\lower0.9ex\hbox{$\sim$}}}
\def \gtrsim{\mathrel{>\kern-1.0em\lower0.9ex\hbox{$\sim$}}}

\documentstyle[11pt,aasms,tighten,flushrt]{article}

\begin{document}
\small

\setcounter{page}{1}

\begin{center}
\bf
DEPARTURE FROM AXISYMMETRY IN PLANETARY NEBULAE
\end{center}

\begin{center}
Noam Soker\\
Department of Physics, University of Haifa at Oranim\\
Oranim, Tivon 36006, ISRAEL \\
soker@physics.technion.ac.il \\
and\\
Saul Rappaport\\
Physics Department, MIT \\
Cambridge, MA 02139 \\
sar@space.mit.edu;
\end{center}

$$
$$

\begin{center}
\bf ABSTRACT
\end{center}

Many planetary nebulae (PNe) exhibit symmetries which range from
unremarkable spherical and elliptical shapes, to quite exotic bipolar
and point-symmetric shapes. However, there are many which exhibit distinctly
non-axisymmetric structure in either (i) the shape of the nebula, or
(ii) in the off-centered position of the illuminating star.
 By examining a large number of well resolved images of PNe we
estimate that $\sim30-50\%$ of all PNe exhibit distinctly non-
axisymmetric structure. In this paper, we discuss how such departures from
axisymmetry can arise from the binary nature of the progenitors of the PNe.
The scenarios include (a) relatively close binaries with eccentric orbits,
and (b) longer orbital period systems with either circular or eccentric
orbits.
In the first mechanism (a), the departure from axisymmetry is caused by the
variation of mass loss and/or mass transfer with the changing distance between
the companions in their eccentric orbit. In the second mechanism (b), the
departure from axisymmetry is the result of the time varying vector direction
of the mass-losing star, or that of a possible pair of jets from
the companion, as the stars move around their orbit.
In order to assess the fraction of PNe
whose non-axisymmetric morphologies are expected to arise in binary systems,
we have carried out a detailed population synthesis study.
In this study, a large number of primordial binaries are evolved
through the lifetimes of both stars, including wind mass loss.
We then assess whether the primary or the secondary (or both)
produces a PN. The expected deviations from axisymmetry are
then classified for each binary and the results tabulated.
We find that $\sim25\%$ of elliptical and $\sim30-50\%$
of bipolar PNe are expected to acquire non-axisymmetric
structure from binary interactions.

{\it Subject headings:} planetary nebulae: general
$-$ stars: binaries
$-$ stars: AGB and post-AGB
$-$ stars: mass loss
$-$ ISM: general


\section{INTRODUCTION}

 Intermediate mass stars form planetary nebulae (PNe) in their
transition from asymptotic giant branch (AGB) stars to
white dwarfs.
 Although single AGB stars are expected to rotate very slowly,
and indeed the extended circumstellar envelopes of most AGB stars
appear spherically symmetric (e.g., Sahai \& Bieging 1993),
most PNe have a large scale axisymmetric rather than a spherical structure
(for recent papers on the subject see Kastner, Soker \& Rappaport
2000).
 The nonspherical PNe can be divided into two main classes,
elliptical PNe and bipolar PNe.
  Bipolar (also called ``bilobal'' and ``butterfly'') planetary nebulae
(PNe) are defined as axially symmetric PNe having two lobes with
an `equatorial' waist between them.
 The elliptical PNe have a large scale elliptically shaped shell,
with no, or only small, lobes or waists.
 These two main classes can be further subclassified according to other
morphological features (e.g., Manchado {\it et al.} 2000).
A few examples of these subclasses are:
($i$)  Inner regions which are axisymmetric, e.g., elliptical, with the
outer regions (mainly the halos), being spherical.
(By ``axisymmetric'' we mean that there exists at least one
axis in 3-dimensions about which the nebula is rotationally symmetric.)
($ii$) More than one axisymmetric substructure,
where the symmetry axes of the different substructures have different
directions, although all symmetry axes basically still pass through
the central star.
 These are termed ``point-symmetric PNe'' (for a recent review
see Manchado {\it et al.} 2000), or ``quadrupolar PNe''
(Manchado, Stanghellini \& Guerrero 1996).
 The common view is that these PNe are formed by precessing jets
(e.g., Livio 2000).
The jets may be blown by the AGB star or post-AGB progenitor,
or from an accreting stellar companion (Soker \& Rappaport 2000, hereafter
SR00).
($iii$) Substructures which have no symmetry axes,
or the symmetry axes of different substructures either lack
a common intersection point or the illuminating star is displaced
from the intersection point.
 These PNe (iii), among other types, are defined by us as having a
{\it departure} from axisymmetry.
 By {\it departure} we refer only to large scale structures, and
not to small blobs, filaments, bubbles etc.

 All classes of PNe may possess departure from axisymmetry.
 Circular PNe possess departure, if their central star is not
at the center of the nebula, and/or they contain a spiral structure
(Soker 1994; Mastrodemos \& Morris 1999).
 The presence of a spiral structure applies also to elliptical
and bipolar PNe.
 In most cases the spiral structure is expected to be smeared
shortly after the ionization by the central star.
Hence, the spiral structure may reveal itself only in the proto-PN
phase.
Bipolar or elliptical PNe would also be said to possess departure
from axisymmetry
if their central star does not lie on their symmetry axis,
or if there is a large scale asymmetry between their two ``sides'';
the latter terminology can refer to either two opposing sides of the
symmetry axis and/or the two sides of the equatorial plane.
 It should be noted that point-symmetric PNe also depart from pure
axisymmetry, but they are not defined by us as having {\it departure}
if they can be built by pure rotations of the different symmetry axes of
the different substructures.
 Only if displacements of one or more of the symmetry axes
relative to the central star is detected, and/or large scale asymmetry
exists, would we define a point-symmetric PN as having a {\it departure}
from axisymmetry.

The axisymmetric structures of PNe have resulted in a long-standing
debate as to whether a binary companion, either stellar or substellar,
is necessary for the formation of these PNe.
 In the present paper we consider another possible role
that binary stellar companions can play, and this is the formation
of PNe with {\it departure} from axisymmetry.

 Four main processes can result in large-scale deviations from axisymmetry.
\newline
{\bf 1. Interaction with the ISM.}
In this case the most prominent features are on the outskirts of the
nebula (e.g., Tweedy \& Kwitter 1994, 1996; Rauch {\it et al.} 2000),
with smaller, or no deviations from axisymmetry in the
inner regions of the nebula.
 In large PNe the ISM may penetrate the outer regions of the nebula, and
influence the inner structure as well as the outer regions
(Dgani \& Soker 1998).
\newline
{\bf 2. Local mass loss events.}
 If one or a few long-lived cool spots exist on the surface of the
AGB star during the mass-loss process itself, they can lead to enhanced
mass loss rates in one to a few particular directions.
 This process seems to be important in massive stars
(e.g., as suggested for the $\sim30 M_\odot$ star HD 179821
by Jura \& Werner 1999), but it is not clear if this process can operate
efficiently in AGB stars, where strong convection may
not allow such spots to live long enough.
\newline
{\bf 3. A close binary companion in an eccentric orbit.}
 This occurs when the companion is close enough to influence the
mass-loss process from the AGB star and/or from the system as a whole,
and the eccentricity is substantial
(Soker, Rappaport, \& Harpaz 1998, hereafter SRH).
 If the companion is close enough to significantly influence the
mass-loss process then a bipolar or elliptical PN is formed.
 In these cases, the companion may (i) tidally spin-up the AGB star
(Soker 1997), (ii) stop the AGB wind if it strongly influences the
wind's acceleration zone (Harpaz, Rappaport, \& Soker 1997),
(iii) accrete and blow a collimated fast wind (CFW;
Morris 1987; SR00), or (iv) gravitationally influence the AGB
wind (Mastrodemos \& Morris 1999).
 If, as we assume here (see also SRH), any of these mechanisms influences
the mass loss rate in a way which depends on the orbital separation,
then the mass loss rate and/or geometry will change periodically
around the orbital, and the nebula will have a departure from axisymmetry.
 For example, if the mass loss is completely stopped at periastron,
then the nebula will have a center of mass velocity, relative to the binary
system, in the direction of motion of the AGB star during apastron passages
(SRH).
\newline
{\bf 4. A wide binary companion.}
 This applies to the case where the AGB star has a wide binary companion
(Soker 1994). Here the departure from axisymmetry results simply from
the fact that the mass loss from the AGB star occurs while it is moving in its
orbit.  Interesting effects due to the orbital motion occur for a wide range
of orbital periods (Soker 1994, Mastrodemos \& Morris 1999).
Soker (1999) takes the condition on the orbital period to be
$0.3 \tau_f \lesssim P \lesssim 30 \tau_f$, where
$\tau_f$ is the formation time of the relevant part of the nebula.
In $\S 2$ we use a somewhat different condition.
 The formation time can be $\tau_f \sim {\rm several} \times 10^4 \yr$
for a PN halo,  $\tau_f \sim {\rm several} \times 10^3 \yr$ for the
dense inner shell, and $\tau_f \simeq {\rm several} \times 100 \yr$
for possible jets.
  Another requirement is that the velocity of the mass-losing star
around the center of mass $v_1$ not be too small.
 This typically requires a companion of mass $M_2 \gtrsim 0.3 M_\odot$,
depending on the orbital separation.
 We note that in the wide orbit cases considered in this work,
the orbital separation may be too large
to allow the companion to directly influence the mass-loss process.
 Hence, the companion has no role in most of these cases in the
formation of the axisymmetrical structure itself
(besides helping to produce a spiral structure).
According to the binary model, a tertiary and closer stellar or substellar
object is required to spin up the AGB progenitor to form any axisymmetric
structure that is observed.

  In a recent paper Soker (1999) analyzes the structure of PNe which were
surveyed by Ciardullo {\it et al.} (1999) for the presence of resolved
visual wide binary companions of their central stars.
 Ciardullo {\it et al.} (1999) used the HST and found
ten PNe for which they argue in favor of a probable physical association of
the
resolved stellar companion with the central star, while for nine others the
association was less likely.
 Soker (1999) analyzes these 19 PNe, and the rest for which no companions
were found, and demonstrates that the departure, or lack thereof, from
axisymmetry of the PNe is consistent in most cases with Ciardullo {\it et
al.}'s claim for an association, or non-association, of the resolved stars
in the PNe.
Another relevant system is the carbon star TT Cygni and its
thin spherical shell (Olofsson {\it et al.} 2000).
 The shell has a radius of $2.7 \times 10^{17} \cm$, it
expands at a velocity of $\sim12.6 \km \s^{-1}$, and its center is
displaced by $\sim1.3 \times 10^{16} \cm$ from the central star,
TT Cygni.  Olofsson {\it et al.} (2000) claim that this implies a relative
velocity of $\sim0.6 \km \s^{-1}$ between the shell and the central star.
 We would attribute this offset directly to the orbital motion itself.
 As noted by Olofsson {\it et al.} (2000), a binary companion at an
orbital separation of $\sim10^3 \AU$ with a mass of $\sim1 M_\odot$
will cause TT Cygni to have this velocity around the center of mass.
The binary system would then have completed $\sim1/4$ to $\sim1/3$ of
an orbital revolution since the shell ejection, $\sim7 \times 10^3 \yrs$ ago.

 A fifth process due to a very wide binary companion can form a local
signature, but does not influence the overall structure of the PN.
  Such a companion, if it has a strong wind, may blow a small bubble
inside the nebula (Soker 1996).
 The bubble may be used to distinguish between stars located
within the nebula and foreground or background stars.
 Such a bubble, formed by a very wide companion, might be the
``vertical bridge'' observed by Corradi {\it et al.} (1999) near a star
located in Wray 17-1.
 We do not consider these systems in the present work.

  In the present paper we estimate the fraction of PNe that are likely
to acquire non-axisymmetrical structures from binary companions,
specifically via processes (3)-(4) listed above.
 In $\S 2$ we describe the criteria used for each process, and
in $\S 3$ we present the results of our population synthesis and
compare them with observations.
 Our main results are summarized in $\S 4$.

\section{CRITERIA FOR BINARY INTERACTION}

\subsection {Wide Companions}

 Simple estimates suggest that the departure from axisymmetry,
e.g., the dislocation of the central star from the center of the nebula,
will be of the order of $\beta \equiv v_1/v_w$, where $v_w$ is the
expansion velocity of the nebula and $v_1$ is the velocity of the mass
losing star around the center of mass (Soker 1994).
 Often, but not always, the expansion velocity, $v_w$, corresponds
approximately to the wind speed of the AGB star progenitor.
 The value of $v_1$ can be expressed as a function of the orbital
separation and the constituent masses as:
\begin{equation}
v_1 = 3
\left( {a} \over {100 \AU} \right)^{-1/2}
\left( {M_2} \over {M_\odot} \right)
\left( {M} \over {M_\odot} \right)^{-1/2} \km \s^{-1},
\end{equation}
 while the orbital period is given by
\begin{equation}
P_{\rm orb} = 1000
\left( {a} \over {100 \AU} \right)^{3/2}
\left( {M} \over {M_\odot} \right)^{-1/2} \yrs,
\end{equation}
where $M=M_1+M_2$, $M_1$ is the mass of the AGB star,
$M_2$ is the mass of the companion, and $a$ is the orbital separation.
 When the eccentricity is not zero, $a$ is taken to be the semimajor
axis, and the velocity given by equation (1) is some average velocity.

 A simple example is a case where a star in orbit ejects a shell of
matter impulsively, i.e., in a time much shorter than the orbital
period. Later, the dust in the shell can reflect the light of the central
star,
 or still later in the evolution, that same star will reveal its hot core and
ionize the shell. Simple geometry gives the offset of the central star
as:
\begin{equation}
   f = \beta \phi^{-1}
\left[ (1-\cos \phi)^2 + (\phi- \sin \phi)^2 \right]^{1/2},
\end{equation}
where $f$ is the distance from the star to the center of the shell in units
of the shell's radius $R_{\rm sh} = v_w \tau_s$, $\tau_s$ is the
age of the shell, and $\phi$ is the angle that the binary has gone through
(in radians) since the ejection event.
  The function describing $f$ has some interesting properties.
  It peaks at a time corresponding to $\sim4$ radians when the maximum
fractional offset is $\sim1.26 \beta$, and asymptotically approaches $\beta$.
 However, it does not reach a value of $0.5 \beta$ until the orbit has
progressed through $\sim1$ radian.
 Therefore, if a $ 5 \%$ effect is required for a case with
$\beta=0.1$, the system should go through at least $\sim1/6$ of an orbit.

 The mass loss during the AGB phase is not impulsive, but is
rather a more continuous wind mass-loss process.
 However, during the final stages of the AGB (as well as in thermal
pulses) there are expected to be strong variations in the mass loss rate.
In such cases, equation (3) may be relevant in describing the ``departure"
from axisymmetry of the corresponding portions of the resultant PN.  For
this to hold, the typical timescales for these variations in the mass
loss rate should be not much longer than the orbital period.
 If the variation timescale is longer, then a spiral structure will
be formed (Soker 1994; Mastrodemos \& Morris 1999).
 If not too many spiral turns have developed, we then also expect to find
non-axisymmetric structure in the PN.  However, in this case such structure
may exist only until the central ionizing source has fully turned on, and
the subsequent heating effects tend to smooth out the ring-like structure.

 In the present paper we assume that any departure of more than
$\sim5 \%$ can be detected,
whether it is a displacement of the central star due to an impulsive
mass loss episode, or non-axisymmetric density structure which leads to a
spiral pattern.
 We therefore simply require that for the detection of a departure
from axisymmetry the velocity ratio should be
\begin{eqnarray}
\beta \equiv {{v_1} \over {v_w}}  > 0.05.
\end{eqnarray}
 Substituting typical values, this condition reads
\begin{eqnarray}
\beta = 0.05
\left( {{a} \over {800 \AU}} \right)^{-1/2}
\left( {{v_w} \over {15 \km \sec}} \right)^{-1}
\left( {{M_1+M_2} \over {2 M_\odot}} \right)^{-1/2}
\left( {{M_2} \over {1 M_\odot}} \right) > 0.05.
\end{eqnarray}
 Note that we have used a characteristic orbital speed rather than
the velocity at periastron or apastron,
since we cannot tell at what position the mass loss
episodes will take place.

 Our second condition is that the orbital period not be
too long or too short.
 As noted from equation (3), for $\beta=0.1$ a departure of $5 \%$
requires $\sim1/6$ of an orbit to be completed.
 We require, therefore, that the binary system completes at least
$1/6$ of an orbit since the relevant mass loss episode occurred, so that
the mass losing star has sufficient time to depart from the center
of the shell.
This gives an upper limit on the orbital period.
 The lower limit on the orbital period is based on
numerical results (Mastrodemos \& Morris 1999; our unpublished results),
and it has to do with the spiral structure that is produced by the
orbital motion when the mass loss rate is continuous rather than
impulsive (Soker 1994).
 If the orbital period is too short, the tight spiral structure will
be smeared very quickly as the nebula is ionized.
Moreover, we have found from our unpublished numerical simulations
that there is a gas dynamical smearing of the rings in only
$\sim6-8$ turns.
 We therefore require that there be no more than $\sim6$ spiral loops
during the appropriate formation time.
 The number of spiral loops is
$N_{\rm spiral} = \tau_f / P_{\rm orb}$, and is limited by the
above arguments to be in the range of
\begin{eqnarray}
1/6 < {{\tau_f}\over {P_{\rm orb}}} < 6 ,
\end{eqnarray}
where $\tau_f$ is the formation time of the particular PN
component under consideration.
 As mentioned in the previous section,
the formation time can be $\tau_f \sim {\rm several} \times 10^4 \yr$
for a PN halo,  $\tau_f \sim {\rm several} \times 10^3 \yr$ for the
dense inner shell, and $\tau_f \simeq {\rm several} \times 100 \yr$
for possible jets.
 Specific limits are set in $\S 3$, where we carry out the population
synthesis study.

\subsection {A close companion in an eccentric orbit}

 SRH demonstrate how a close companion in an eccentric orbit can cause
the central star to be displaced from the center of the nebula.
 They postulate that the mass loss rate from the AGB star varies
systematically with orbital phase, hence the center of mass
velocity of the nebula will be different than that of the
center of mass of the binary system.
 SRH consider several effects, including a tidal enhancement of the
stellar wind near periastron, and a cessation of the stellar wind when
the Roche lobe of the AGB star encroaches on its extended atmosphere
near periastron passage.
 Tidal effects require the companion to come close to the
AGB star at periastron passages. SRH find that the
condition for this is $0.4 R_L \leq R_g \lesssim R_L$,
where $R_L$ is the critical potential lobe of the AGB star at periastron,
and $R_g$ is the AGB stellar radius.
 We find that these close systems have strong tidal interactions
and hence reach circularization, i.e., e=0.
 These will not cause any departure from axisymmetry by our criteria.
 The condition used by SRH for the cessation of the AGB wind to be
significant is that at periastron passage
\begin{eqnarray}
R_{\rm wfz} \gtrsim R_L,
\end{eqnarray}
where $R_{\rm wfz}$ is the radius of the ``wind formation zone''
(wfz), i.e., the region from which the AGB wind is accelerated.
 Harpaz {\it et al.} (1997) consider a wind's acceleration zone
to extend to $R_{\rm wfz} = 10 R_g$.
 We here use a more ``conservative'' approach, and
take $R_{\rm wfz} = 5 R_g$.
 Another condition to be fulfilled is that the eccentricity be
$e \gtrsim 0.2$, and that the companion mass be above
some minimum value, which we take to be $0.5 M_\odot$.
  The SRH results may pertain to binary systems with semimajor axes in the
range of  $a \simeq 7$ AU to 80 AU, which correspond
to  orbital periods in the range  of $ P \simeq 15 - 500$ years.
 SRH propose that this mechanism can apply to the
bipolar PN MyCn 18 (the Hourglass Nebula), where the central star is
displaced from the center of the nebula (Sahai {\it et al.} 1999).

 In the present paper we consider another process which may cause
departure from axisymmetry in relatively close eccentric binary systems.
 This is the formation of a collimated fast wind (CFW) by the companion
(Morris 1987; SR00).
 The companion is assumed to accrete from the AGB wind, to form an accretion
 disk, and to blow a CFW.
  The interaction between the CFW, if strong enough (for exact condition
see SR00), and the AGB wind will form a bipolar PN (Morris 1987).
 If the companion has an eccentric orbit, then the mass accretion
rate, and hence the CFW's strength by our assumption, will change around
its orbit. If the CFW is strong enough and the periodic changes in its
intensity are large enough, this may lead to a departure from axisymmetry
because both the orbital velocity of the star blowing the CFW and
the interaction pattern between the CFW and the slow AGB wind
will change with orbital phase.
 The changes in accretion rate and orbital velocity were
mentioned briefly by Miranda {\it et al.} (2001b, their $\S 3.5$)
as a possible effect in the PN Hu 2-1, and was also discussed in a
theoretical context by Soker (2001b).  Obviously, detailed hydrodynamic
simulations are required to understand these effects more quantitatively.

 We assume that this CFW-mechanism for producing departure from
axisymmetry is important whenever the following conditions are met.
  First, the accretion rate at periastron $\dot M_2$ is such that
\begin{eqnarray}
\dot M_2 > \mu  \vert \dot M_1 \vert  ,
\end{eqnarray}
where $\dot M_1$ is the mass loss rate from the AGB star, and
$\mu \sim0.01-0.1$.
Second, the mass accretion rate at apastron should be lower than at
periastron by a
factor of $\gtrsim 2$. The last condition implies that $e \gtrsim 0.2$.

 If circularization of the binary occurs, the processes
discussed in this subsection will not be important.
 In our population synthesis study (see Sect. 4) we check for
circularization as in SR00.
 We note, however, that there are close binary systems with $a \simeq 1 \AU$
that {\it do} have eccentric orbits (Van Winckel 1999;
Van Winckel, Waelkens, \& Waters 1995) even though our simple
formula would indicate that circularization should have occurred.
 In our formulation, we may therefore miss some interesting cases.

\section{POPULATION SYNTHESIS}

In our population synthesis and evolution study, we utilize Monte
Carlo techniques, and follow the evolution of some $5 \times 10^4$
primordial binaries.  For each primordial binary, the mass of the primary
is chosen from an initial mass function (IMF), the mass of the secondary
is picked according to an assumed distribution of mass ratios for primordial
binaries, the orbital period is chosen from a distribution covering all
plausible periods, and the orbital eccentricity, $e$, is chosen from a uniform
distribution (the details of all these prescriptions and procedures are
given in SR00).   Once the parameters of the primordial binary have been
selected, the two stars are evolved simultaneously using relatively simple
prescriptions (SR00).  We explicitly follow the wind mass loss
of both stars, at every step in the evolution.  For this purpose we have
developed a wind mass loss prescription that depends on the mass and
evolutionary state of the star, and that is designed to reproduce
reasonably well the observed initial-final mass relation for single stars
evolving to white dwarfs (see SR00 for details).  We also take into
account the evolution of the binary system under the influence of
stellar wind mass losses.

        At each step in the evolution, we compute the fraction of the
stellar wind of one star that will be captured via the Bondi-Hoyle
accretion process by its companion.  In addition to the mass capture rate,
we also estimate whether sufficient angular momentum will be accreted to
allow for the formation of an accretion disk before the accreted matter
falls on the companion, and we check whether the total rate of accretion
exceeds a certain critical value to form a CFW (see SR00;
note that an exponent of $1/2$ is missing in the second parenthesis
of SR00's equation [2] for the condition on accreted angular momentum).
Finally, we test whether tidal forces will circularize the binary
before the onset of the superwind (the final intensive wind [FIW]
at the end of the AGB) phase.

Unlike the study of SR00 we are also very much interested in {\it wider}
binary systems where there is little or no interaction between the AGB star
wind and the companion.  We therefore also keep track of these systems
and their properties during the evolution.

We have added to the population synthesis code new segments to
specifically examine the fraction of systems with the specified binary
parameters that were outlined in the previous section and are enumerated
in Tables 1-3 below.

\subsection{Results}

 In Tables 1-3 we summarize the number of PNe that are expected to exhibit
departure from axisymmetry via the influence of  wide binary companions
and/or close eccentric binary systems, under the prescribed constraints.
The meanings of the different symbols in Table 1-3 are as follows:
$v_{\rm min}$ is the minimum allowed orbital velocity of the mass losing star
around the center of mass, $v_1$ (utilized in the constraint given by
equation [4]);
$\tau_f$ is the formation time of the relevant component in the PN, and is
used in the constraint on the orbital period as given by equation (6);
$e_{\rm min}$ is the minimum allowed orbital eccentricity;
and $\mu$ is the minimum ratio of the Bondi-Hoyle accretion rate to
the mass loss rate of the AGB star which is used in the constraint given
by equation (8).
The classes are as indicated in the Tables, where A--E are wide binary cases,
and G--K are close eccentric binary systems.
 For class K, which is the SRH mechanism, the conditions are:
($i$) according to equation (7),
where the radius of the wind's acceleration region is set to
$R_{\rm wfz} = 5 R_g$, where $R_g$ is the AGB stellar radius,
and we set $R_L = 0.38 (1-e) a$ at periastron;
($ii$) the companion mass is $>0.5 M_\odot$; and ($iii$) $e>0.2$.

The results are expressed in Tables 1--3 as a percentage of the total
number of binary systems we start with, which is about equal to the
total number of systems which form PNe (see SR00 for more details
regarding the efficiency of forming PNe in binary systems).
 Each tabulated value indicates how many systems belong both to the
 class along the row and the class along the column.
 For example, $5.1 \pm 0.1 \%$ of all binary systems have the properties
of class A, while $4.0 \pm 0.1 \%$ have both the properties of class A
and class B (Table 3).
  Table 1 is for systems where the initially more massive star forms
the PN, while Table 2 is for systems where the initially less massive
star is the progenitor of the PN.
Table 3 is the sum of Table 1 and Table 2.
 The indicated uncertainties are statistical, taken as the square
root of the number of systems in each group.
 However, the uncertainties involved in the criteria for detecting
departure from axisymmetry are much larger.
 Hence, the number in the table are given to the accuracy of two
significant digits.

 We note that the different classes have the following observational
consequences:
 {\bf Class A:} PNe with departure from axisymmetry in the halo, allowing
for very small degrees of departure ($v_w=15 \km \s^{-1}$ and
$\beta >0.05$ in eq. 4). {\bf Class D:} PNe with a large
degree of departure from axisymmetry in the halo ($\beta>0.15$).
 {\bf Class B:} departure from axisymmetry of the main PN shell,
allowing for very small degrees of departure. {\bf Class E:}
PNe with a large degree of departure from axisymmetry in the main shell.
 {\bf Class C:} departure from axisymmetry is expected from jets, if they are
formed.  Here we assume the typical velocity of the jets to be
$v_j = 100 \km s^{-1}$, and take $\beta>0.05$.
 Classes A--E refer to the influence of a wide companion,
whether in an eccentric or circular orbit.
 Classes G--K are closer systems, where a strong interaction with
the companion star, including possible accretion, is important.
Many of the systems in classes G-K will form bipolar PNe (SR00).
In cases where the polar outflows are very fast, the departure from
axisymmetry may be noticed only in the slower equatorial flow.
 Such a case might be the Egg nebula (CRL 2688), with its two highly
symmetric ``search-light'' beams (Sahai {\it et al.} 1998a, b).
 In this object, the equatorial disk, as revealed by the NICMOS camera,
shows a clear departure from axisymmetry (Sahai {\it et al.} 1998a,b).

 The main findings from the population synthesis are as follows:
\newline
(1){\it $\sim5 \%$ of halos should show departure from axisymmetry.}
Of all PNe which have a halo, $\sim5 \%$
are expected to show signs of departure from
axisymmetry due to a wide binary companion (class A).
 Only an extremely small fraction ($\sim0.01 \%$)
will have a large departure (class D).
 Approximately 3/4 of class A systems, or $4 \%$ of all PNe,
will show a departure in their main shell as well (classes B and A).
 We note also that in $\sim1/20$ of PNe which possess a non-axisymmetric
halo ($0.23 \%$ of all PNe), the companion may have influenced
the structure due to its eccentric orbit as well (classes A and I).
 These are systems with large eccentricity, so that the companion gets
close to the mass losing star at periastron passage.
 The main problem with PNe belonging only to class A, i.e., only
the halo shows a departure from axisymmetry is how to distinguish
between departure caused by a wide binary companion
and that caused by interaction with the ISM, as the large tenuous
halos are expected to be significantly influenced by the ISM.
\newline
(2) {\it $\sim25 \%$ of elliptical PNe should exhibit departure from
axisymmetry.}
   Only $\sim1.2 \%$ of all PNe with a halo (and not many PNe
have an observable halo) should have departure only in their halo.
This number is estimated from class A systems ($5.1 \%$) minus
systems belonging to class A and any other class, e.g.,
classes A+B have $4.0 \%$ of all PNe with s halo.
  Only $\sim1.3\%$ may have departure from their jets, if they exist
(class C alone).
 Therefore, the majority of PNe that have a significant departure from
axisymmetry due to a wide companion come from class B, and they amount to
$\sim17 \%$. Most of these will be elliptical or spherical PNe.
 If we add several more percent of elliptical PNe that result from the
influence of eccentric companions at large distances (mainly from
class G), and remember that $\sim90 \%$ of all PNe are elliptical
or spherical PNe, we conclude that $\sim25 \%$ of all elliptical
PNe have an observable departure from axisymmetric due to a companion.
\newline
(3) {\it We underestimate PNe with departure from axisymmetry because
of strong tidal interactions.}
In our simulation we check for tidal circularization
(see SR00 for details).
 The eccentricity of systems for which the circularization time is shorter
than the evolutionary time is set to $e=0$, hence they cannot
cause departure from axisymmetry.
 However, it is well known that there are binary systems with post-AGB
stars, which have orbital periods ranging from less than a year up to a
few years, and substantial eccentricities  $0.1 \lesssim e \lesssim 0.4$
(e.g., Van Winckel 1999; Waelkens {\it et al.} 1996).
  Such close binary systems, with orbital separations
of a few AU, will cause departure from axisymmetry via the
SRH mechanism (see $\S 2.2$), in which the companion directly influences
the mass loss process (as in class K), as well as by accreting from the
mass losing star (as in classes G-J discussed above).
 Therefore, as noted already in $\S 2.2$, we may miss some bipolar PNe
which do have departure from axisymmetry due to a companion
in an eccentric orbit where the Roche lobe remains well outside the AGB
envelope.
\newline
(4) {\it We underestimate PNe with departure from axisymmetry because
of Roche lobe overflow.}
 We do not simulate systems that go through Roche-lobe
overflow or a common envelope. However, systems which go through a
common envelope may still possess substantial departure from axisymmetry.
This is evident from the bipolar PN NGC 2346 which exhibits considerable
departure from axisymmetry in its equatorial plane (e.g., see Corradi \&
Schwarz 1995), and has a binary central star which went through
a common envelope (Bond \& Livio 1990).
\newline
(5) {\it $\sim30-50 \%$ of bipolar PNe should show departure from
axisymmetry.}
 The fraction of systems that belong to class G but do not have
too long of an orbital period, i.e., do not belong to class B+G, is
$\sim5 \%$.
 To these we add $\sim1 \%$ that belong to class K, but not
to classes B or G.
 We find, therefore, that $\sim6 \%$ of all PNe have departure and
have close companions.
 Most PNe that belong to classes G-K above will form bipolar PNe
(but not all of them).
 So not all of the $\sim6 \%$ will form bipolar PNe, but somewhat
less than that, $3-5 \%$.
 Now, since $\sim10 \%$ of all PNe are expected to be bipolar according
to the binary model of SR00, and from observations this
fraction can be somewhat larger (Manchado {\it et al.} 2000), we
estimate that $\sim30-50 \%$ of all bipolar PNe may show
departure from axisymmetry.
 This number is a crude estimate since as noted in point 3 and 4 above,
many bipolar PNe are expected to be formed by systems not simulated here,
e.g., those with Roche lobe overflow (see SR00), and some of
these may possess departure from axisymmetry.
  Considering these arguments, we can safely say that $\sim30-50 \%$ of
all bipolar PNe will show a departure from axisymmetry.
  Since the lobes of many PNe move at high velocities, the departure
from axisymmetry will be easier to detect in the slow equatorial flow
(e.g., the Egg nebula mentioned above).
 The point-symmetric structure of many PNe, especially bipolar PNe,
will make the detection of departure from axisymmetry more
difficult.

 The percentages of elliptical and bipolar PNe that are expected to
possess departure from axisymmetry due to a binary companion, according
to the population synthesis results are summarized in Table 4
(rows labeled ``Theory'').

\subsection{Comparison with Observations}

 This subsection is not meant to present a rigorous statistical analysis;
we do not use any complete samples of PN observations, and we do not
conduct a thorough analysis of the different signatures of departures
from axisymmetry.
 Our sole purpose here is to show that the estimates found from the
population synthesis are compatible with available
observations. The results are summarized in Table 4.

\subsubsection  {Elliptical PNe}
 In this subsection we evaluate the empirical evidence for departure
from axisymmetry of elliptical PNe by direct inspection of three different
available PNe data sets.
First, Soker (1997) used a data set of all PNe with resolved images
which were available to him to examine morphologies in the context of
binary models for the shaping of PNe.  Many of these images were of too
low a quality to permit detection of small degrees of departure from
axisymmetry. He classified the PNe according
to the type of binary interaction that may have shaped them:
($i$) PNe whose progenitors did not interact with any close companion
(these form spherical PNe);
($ii$) PNe whose progenitors interacted with close stellar companions
outside their envelopes (these form bipolar or elliptical PNe
[SR00; Soker 2001b]);
($iii$) PNe whose progenitors interacted with a stellar companion via
a common envelope phase (these form bipolar or extreme elliptical PNe);
and ($iv$) PNe whose progenitors interacted with substellar objects,
but not with a close stellar object (these form moderate elliptical PNe).
 Soker (1997) estimated that out of 293 PNe which did {\it not} interact
with a close stellar companion (his tables 2 and 5; classes $i$ and
$iv$ above), the structure of $\sim25$ was influenced by a wide
binary companion (PNe marked WB or ISM/WB in his tables; ISM/WB means an
interaction with the ISM may be an alternative
explanation to the presence of a wide binary companion).
 From 113 PNe that presumably interacted with a stellar companion via
a common envelope evolution (Soker 1997; his Table 4), Soker finds
$10 \pm 4$ to show some signature of departure from axisymmetry due to
an interaction with a companion.
 These findings mean that the structure of $\sim8-9 \%$ of elliptical
PNe are likely to show signatures of departure from axisymmetry
due to a wide companion, according to the
``conservative'' approach used by Soker (1997).
 This result is summarized in Table 4 (of the present paper; see
row labeled ``Soker 1997'').
 Soker did not consider small departures, and therefore avoided
the problem of large clumps and filaments, but on the other hand he missed
many PNe which do possess departure due to a binary companion.
 Therefore, Soker (1997) underestimated the number of PNe in which structure
has been influenced by a wide companion.
 Soker's (1997) numbers should be compared with PNe expected to
possess a large departure from axisymmetry, i.e., Class D+E ($1.9 \%$ in
Table 3), and a fraction of Class H, that which has an overlap with
class A-D, but not with Class E ($0.2 \%$ in Table 3).
In total, $\sim2.1 \%$ of all PNe, or $\sim2.5 \%$ of elliptical PNe.
 The fraction found by Soker (1997) is between those expected to possess
large departure and the total number we expect to possess detectable
departure. We consider this a satisfactory result.

 Another search for departure from axisymmetry in a large sample of
PNe was conducted by Soker (1999), who studied the sample of
Ciardullo {\it et al.} (1999).
 Using the HST, Ciardullo {\it et al.} (1999) surveyed 113 PNe for
 the presence of resolved visual wide binary companions of their
central stars.
 For 19 PNe they  argue for probable or possible association of the
resolved stellar companions with the central stars.
 Soker (1999) found that for $60 \%$ of the elliptical PNe among these
19 PNe, the departure likely resulted from a wide companion
(the fraction is $\sim80 \%$ for the 10 PNe for which Ciardullo
{\it et al.} argue for a probable association).
 Soker (1999) found that for the rest of the PNe in their sample, the
fraction with observable departure from axisymmetry due to a wide
companion is $\sim35 \%$.
 Overall, the total fraction of elliptical PNe showing departure
due to a wide companion in the list of PNe surveyed by Ciardullo
{\it et al.} (1999) is estimated to be $\sim40 \%$
(see Table 4; row labeled ``Soker 1999'').
This is somewhat higher than the results of the population synthesis
which indicates only $\sim25 \%$.

 To study small departures from axisymmetry we examined high
resolution HST images of PNe.
 We examined 71 images (same list as the one collected by
Terzian \& Hajian 2000), for which the different HST images can be
found in the following papers:
Balick (2000), Balick {\it et al.} (1998), Bobrowsky {\it et al.} (1998),
Bond (2000), Borkowski, Blondin, \& Harrington (1997),
Corradi {\it et al.} (2000), Kwok, Su, \& Hrivnak (1998), Sahai (2000a,b),
Sahai \& Trauger {\it et al.} (1998),
Su {\it et al.} (1998), and Terzian \& Hajian (2000).
 A more accurate analysis including a detailed list of the PNe
is postponed to a future project.
 We count only PNe for which the departure from axisymmetry can be clearly
discerned in these images.
  Many of the images show large clumps and filaments which result
from the stochastic nature of the mass loss process, rather than
from a companion.
 Examples are NGC 6210 and NGC 6326 which we still count
here as having departure, although it is very likely that these are
due to the stochastic nature of the mass loss process, or other
instabilities.
 Therefore, we here overestimate the number of PNe which acquire their
departure from axisymmetry from a companion.
  Of these 71 images, 4 are difficult to classify based on
the images alone, while 26 are bipolar PNe, and 41 are elliptical PNe.
 Out of the 41 elliptical PNe, many have large clumps and/or filaments
in their inner regions, making it very difficult to decide whether
a departure from an axisymmetric structure due to a companion exists.
 For 12 PNe we could not tell whether they do or do not possess departure
from axisymmetry, 16 PNe do show a departure,
while 13 do not exhibit a clear departure from axisymmetry.
  These results alone indicate that $\sim50 \%$ of elliptical PNe
show some signature of departure from axisymmetry
(see Table 4; row labeled ``HST images'').
 Again, many are due to clumps and filaments which result from the
stochastic nature of the mass loss process.

  Combining the three different estimates (Soker 1997, 1999,
and the study of the 71 HST images), as summarized in Table 4,
we argue that our estimate from the
population synthesis that $\sim25 \%$ of all elliptical PNe should have
a detectable departure from axisymmetry acquired from a companion
is compatible with observations, although it seems that the fraction
of observed elliptical PNe with departure is somewhat higher
than $25 \%$, and is $\sim35 \%$.
  This modest discrepancy can be accounted for by three effects:
($i$) Binary systems with longer or shorter periods, or with
slower orbital motion than can satisfy our selection criteria can
still cause a noticeable departure from axisymmetry
(see Soker 2001b for formation of CFW in somewhat wider binary systems).
($ii$) Many PNe acquired their departure from axisymmetry from
the stochastic nature of the mass loss process.
 It is likely that the stochastic nature will be more
prominent in the equatorial plane, making it even more difficult to
distinguish between the mass loss process and a wide companion.
($iii$) The number of PNe which acquired their departure from axisymmetry
via interaction with the ISM may be larger than what has been estimated.
 This latter mechanism has a small effect since we avoided PNe
showing departure only in their very outer regions (halos).
 We could not avoid including some PNe which acquired their departure
via mechanism ($ii$).
However, many PNe show the same sense of departure in two or more
regions. A stochastic process cannot explain these cases;
hence, it will not account for all, or even most, of the cases with departure
we consider here.

\subsubsection {Bipolar PNe}
 Examining the 43 images of bipolar PNe given by Corradi \& Schwarz
(1995), we  find that many show a clear departure from axisymmetry,
mainly in the equatorial plane.
Examples are 19W32; NGC 2899; NGC 650-1; Sh 1-89; We 1-4.
However, many require very careful examination.
 Such is the case with MyCn 18, whose lobes seem quite axisymmetric,
but the central star is not in the center of the inner region as revealed
by the HST (Sahai {\it et al.} 1999).

 We turn to the 71 HST images mentioned in the previous subsection,
noting that even with these high resolution observations we may still
miss some PNe which do possess departure from axisymmetry, since
departure has been revealed, or may reveal itself in other bands of the
spectrum.
For example, the Red Rectangle (BD-10-1476) contains a close
eccentric binary system, with $P_{\rm orb} = 318$ days and $e=0.38$
(Waters {\it et al.} 1998).
The general structure of the Red Rectangle is highly axisymmetrical, up
to a distance of $\sim1 ^\prime$ from the central star
(e.g., Van Winckel 2000).
However, the 10 $\mu$m map presented by Waters {\it et al.}
(1998; their fig. 3) shows a clear departure from axisymmetry on scales
of $\sim5 ^{\prime \prime}$ from the central star.
 Their contour map shows that the equatorial matter is more
extended on the west side.
 Another example is the Egg nebula which appears axisymmetric in the HST
optical image, while only the NICMOS
image shows a clear departure from axisymmetry (Sahai {\it et al.} 1998a,b).

 From the 26 bipolar PNe in the list of 71 HST images, 9 possess a clear
departure from axisymmetry, 11 do not, and for 6 PNe it is difficult to tell
due to the presence of numerous ``blobs" and/or filaments.
 Based on these images alone, it turns out that $\sim35 - 60 \%$ of all
bipolar PNe possess departure from axisymmetry (Table 4).
 However, we still miss some (e.g., the Egg nebula).
 On the other hand, many PNe acquire their departure from axisymmetry
from large blobs and filaments, as noted above
for elliptical PNe.
An examination of 52 bipolar PNe Soker (1997; his table 3) indicated
departure from axisymmetry in 5 out of 52 PNe, i.e, $\sim10 \%$.
Of these, the departure in 3 ($\sim6 \%$) PNe was attributed to a
companion, and for 2 PNe the departure was attributed to interaction with
the ISM.
We take the percentage found by Soker (1997) to be $\sim8 \%$
(Table 4).
 As noted above, Soker did not consider small scale departure, and
therefore avoided the problem of large clumps and filaments, but on
the other hand he missed most of the PNe which do possess departure.
  The fraction found by Soker (1997) should be compared with
the expected number of PNe which possess large departure, mainly
Class J which has $\sim1 \%$ (Table 3) of all PNe,
or $\sim10\%$ of bipolar-PNe.

 Overall, it seems that our finding from the population synthesis that
$30-50 \%$ of all bipolar PNe acquire their departure from a companion,
is compatible with available observations.

\section{SUMMARY}

 We used a population synthesis code to estimate the number
of PNe expected to possess detectable departure from axisymmetric
structure as a result of a binary companion.
 For that we considered mechanisms for causing departure from axisymmetry
as proposed by us in earlier works;
a wide companion with a long orbital period (Soker 1994),
and/or a close companion in an eccentric orbit (SRH).
 The departure can manifest itself as the illuminating star not being at the
center of the nebula, one side being brighter or more extended than the other,
the two sides having different magnitude Doppler shifts, or any
combination of these.
  Point symmetric structures, where the nebula can be built from
several axisymmetric structures all having the same center for
their symmetry axis, are not considered by us as exhibiting departure
from axisymmetry.
   We set different specific values in the criteria for causing significant
departure from axisymmetry:
orbital period, orbital velocity of the mass losing AGB star around
the center of mass, and formation time of the relevant nebular part
in the case of wide binaries ($\S 2.1$, eqs. 4 and 6),
and eccentricity and mass accretion rate for the close eccentric companions
($\S 2.2$).
These values are indicated in Table 1-3 for the different classes.
A crude comparison with observations is summarized in Table 4.
 Our main findings are:
\newline
 (1) $\sim25 \%$ of all elliptical or circular PNe are expected to
possess detectable departure from axisymmetry
(mainly classes A-E and some fraction of class G in Table 3).
 Of these $\sim25 \%$, in $\sim19 \%$ the initially more massive
star is the AGB mass losing star which has a main sequence companion
(Table 1), and in $\sim6 \%$ the initially less massive star is the
central star and the companion is a WD (Table 2).
\newline
(2)  $\sim30-50 \%$ of all bipolar PNe are expected to
possess detectable departure from axisymmetry (most of the systems
in classes G-K; Table 3).
 In most of these systems the AGB mass losing star is the initially more
massive star and the companion is a main sequence star (Table 1).
 Only $\sim 5-8 \%$ out of these $\sim30-50 \%$ have a
WD companion (Table 2), while the rest, $\sim 25-40 \%$, have a
main-sequence companion.
\newline
(3) Since $\sim10-15 \%$ of all PNe are bipolar, and the
rest are elliptical or circular,  we find from our results
(first row of Table 4), that $\sim 27 \%$ of all PNe are
expected to possess significant departure from axisymmetric structure.
\newline
(4) We find satisfactory agreement in the fraction of PNe possessing
departure from axisymmetry between the results of the
population synthesis and the rough estimate we made from the observations
(Table 4, last two rows).
 The largest uncertainty in the theoretical results are in the
criteria for detectable departure from axisymmetry.
 The main difficulty in analyzing observations is to distinguish
between a departure from axisymmetry caused by a companion, and that
caused by large scale instabilities and stochastic mass loss process.
 Considering these uncertainties, we find the agreement reflected in Table 4
to be satisfactory.

In light of our findings that a large fraction of PNe are expected to
possess observable departure from axisymmetry, more attention should
be paid to this effect when analyzing images and velocity maps of PNe.
 Some recent papers, indeed, do that.
 A departure from axisymmetry, in that the central star is not at
the center of the nebula, is noted by Sahai (2000b) in the two PNe
He 2-47 and M1-37.
 Miranda {\it et al.} (2001b; for Hu 2-1) and Miranda, Guerrero, \&
 Torrelles (2001a; for IC 4846), proposed that the difference between
the systematic velocity of the precessing jets and the centroid velocity
of the nebulae in these two PNe results from the orbital motion of the
star that blows the jets.
This is one of the manifestations of departure from axisymmetry
in binary systems.
Finally, we note that departure from axisymmetry can occur in
other systems similar to PNe.
 Soker (2001a) argues that the departure of the nebula around
$\eta$ Carinae can be explained by the presence of the proposed
binary companion, if it has an eccentric orbit.

\bigskip

{\bf ACKNOWLEDGMENTS:}
 This research was supported in part by grants from the US-Israel
Binational Science Foundation,
and NASA under its Astrophysics Theory Program: Grants NAG5-4057 and
NAG5-8368.


\end{document}